\documentclass[twocolumn,final]{svjour3}
\usepackage{amsmath}
\usepackage{amssymb}
\usepackage{subfigure}
\usepackage{graphicx}% Include figure files
\usepackage{bm}% bold math
\newcommand{\vect}[1]{\boldsymbol{#1}}

\begin{document}
\title{Electronic structure in underdoped cuprates due to the emergence of a pseudogap}
%Evolution of the electronic structure of underdoped cuprates in the YRZ model}

\author{J. P. F. LeBlanc \and J. P. Carbotte}

% \altaffiliation[Also at ]{Physics Department, University of Guelph.}
%\author{J. P. Carbotte$^{3,4}$}
%\author{E. J. Nicol$^{1,2}$}%
\institute{J. P. F. LeBlanc$^{1,2}$ \at $^1$ Department of Physics, University of Guelph,
Guelph, Ontario N1G 2W1 Canada \email{leblanc@physics.uoguelph.ca} \and $^2$Guelph-Waterloo Physics Institute, University of Guelph, Guelph, Ontario N1G 2W1 Canada \and \\
J. P. Carbotte$^{3,4}$ \at $^3$ Department of Physics and Astronomy, McMaster
University, Hamilton, Ontario L8S 4L8 Canada \and $^4$ The Canadian Institute for Advanced Research, Toronto, ON M5G 1Z8 Canada} 
%\address{$^2$Guelph-Waterloo Physics Institute, University of Guelph, Guelph, Ontario N1G 2W1 Canada}
%\address{$^3$ Department of Physics and Astronomy, McMaster
%University, Hamilton, Ontario L8S 4L8 Canada}
%\address{$^4$ The Canadian Institute for Advanced Research, Toronto, ON M5G 1Z8 Canada}
\date{\today}
\journalname{Journal of Superconductivity and Novel Magnetism}
%\author{J.P.F. LeBlanc and J.P. Carbotte}

%\submitto{\JPCM}
\maketitle
\begin{abstract}
The phenomenological Green's function developed in the works of Yang, Rice and Zhang has been very successful in understanding many of the anomalous superconducting properties of the deeply underdoped cuprates.  It is based on considerations of the resonating valence bond spin liquid approximation and is designed to describe the underdoped regime of the cuprates.  Here we emphasize the region of doping, $x$, just below the quantum critical point at which the pseudogap develops.  In addition to Luttinger hole pockets centered around the nodal direction, there are electron pockets near the antinodes which are connected to the hole pockets by gapped bridging contours.  We determine the contours of nearest approach as would be measured in angular resolved photoemission experiments and emphasize signatures of the Fermi surface reconstruction from the large Fermi contour of Fermi liquid theory (which contains $1+x$ hole states) to the Luttinger pocket (which contains $x$ hole states).  We find that the quasiparticle effective mass renormalization increases strongly towards the edge of the Luttinger pockets beyond which it diverges.
   \keywords{Underdoped cuprates \and Mott transition \and Pseudogap \and Fermi surface reconstruction \and Resonating Valence Bond \and ARPES \and Effective Mass}
   \PACS{71.18.+y \and 74.20.Pq \and 74.25.Jb}%74.72.-h, 74.20.Mn, 74.25.Gz, 78.30.-j}
    \end{abstract}
\section{Introduction}
%\setstretch{2}
A first qualitative understanding of the superconducting properties of the cuprates at optimal and overdoping is provided by BCS theory with the important modification that the gap has $d$-wave symmetry which requires nodes and a change in sign on crossing the nodal direction, $(\pi,\pi)$.  In the underdoped regime, however, anomalous properties are observed which require new elements
\cite{yrz:2006,valenzuela:2007,bascones:2008,yrz:2009,guyard:PRL:2008,tacon:2006,loram:1998,anukool:2009,kanigel:2006,kanigel:2007,norman:2007,kondo:2009}  for their understanding, which fall outside of $d$-wave BCS \cite{o'donovan:1995:4568,o'donovan:1996,branch:1995,o'donovan:1995:16208}
and go beyond effects of inelastic scattering,\cite{nicol:1991,schachinger:1997,schachinger:2000,schachinger:2003,carbotte:2005,marsiglio:1996} strong coupling,
\cite{marsiglio:1987,mitrovic:1980,schachinger:1980} and anisotropy.\cite{leung:CJP:1976,tomlinson:1976,leung:1976}
Models that might account for some of the unconventional properties have been proposed which include the possible role of phase fluctuations,\cite{franz:1998} locally fluctuating antiferromagnetic order,\cite{qi:2010} and the existence of preformed pairs.\cite{emery:1995}  These pairs are envisaged to form at a higher temperature, T$^*$ (the pseudogap energy scale), than the superconducting T$_c$ where phase coherence is achieved.  There are also models of competing orders, such as spin  or d-density waves.\cite{chakravarty:2001,zhu:2001}
  Here we consider instead the recent model of Yang Rice and Zhang (YRZ)\cite{yrz:2006} which is based on a microscopic theory for a resonating valence bond (RVB) spin liquid.\cite{lehur:2009}  YRZ provides an ansatz for the coherent part of the charge carrier Green's function.  The proposed self energy, which describes the formation of a pseudogap on the antiferromagnetic Brillouin zone (AFBZ) boundary, for doping, $x$, below a quantum critical point (QCP) at a critical doping of $x=x_c$, has the very desirable property that it is relatively simple and therefore calculations of properties are often straightforward.  Another important element of the theory, in its final form, are the Gutzwiller factors which describe the narrowing of the electronic bands as the Mott transition to an insulating state is approached with reduced doping.  An additional Gutzwiller factor accounts for a reduction in the magnitude of the coherent part of the Green's function as the incoherent part grows. 
  
   Many of the superconducting properties of the underdoped cuprates, which were previously considered to be anomalous, have recently been understood
  \cite{valenzuela:2007,leblanc:2010,leblanc:2009,carbotte:2010,illes:2009,bascones:2008,yrz:2009} qualitatively within the model of YRZ and this encourages us to explore further its predictions.
  The properties considered so far include: Raman scattering\cite{guyard:PRL:2008,tacon:2006} which shows two gap scales,\cite{valenzuela:2007,leblanc:2010} which in the YRZ model are identified as combinations of pseudogap and superconducting gap peaks.  The large reduction in the normalized specific heat jump at T$_c$ as $x$ is decreased \cite{loram:1998} is also easily understood,\cite{leblanc:2009} as is the penetration depth data\cite{anukool:2009} indicating that the slope of the low temperature, T, linear in T law is much less affected by doping than is the magnitude of the zero temperature superfluid density.\cite{carbotte:2010}  
A comparison of Fig.~7(b) of reference \cite{carbotte:2010} with data on Bi-2212 for Tc vs the inverse square of the penetration depth at zero temperature, which is referred to as the Uemura plot, shows good agreement.  A critical element of this success is the appearance  of the square of the Gutzwiller factor, $g_t(x)$, in YRZ which goes like $(\frac{2x}{1+x})^2$.  The data only goes down to transitions temperatures of approximately one half that of the optimal critical temperature.  It may be that to go closer to the end of the superconducting dome in the highly underdoped regime, additional effects not included in YRZ may need to be considered such as phase fluctuations.  
  Hat-like structures in the optical self energy which emerge in the underdoped regime\cite{illes:2009} and are not present at optimum and overdoping are explained as is the second energy scale beyond the superconducting scale which is seen in the partial optical sum.\cite{homes:2004}  The checkerboard pattern\cite{bascones:2008} observed in Fourier transform scanning tunneling spectroscopy (FT-STS)\cite{mcelroy:2003,mcelroy:2005,hanaguri:2007} has been discussed, as have some aspects of the angular resolve photo-emission (ARPES) data\cite{kanigel:2006,kanigel:2007,norman:2007} and its relationship to FT-STS\cite{yrz:2009}.  

An important success of the YRZ model, among many others in addition to those just mentioned, is that the ARPES results of Kondo et al.\cite{kondo:2009} for highly underdoped Bi-2201 are easily understood.\cite{leblanc:2010}  What is measured is the energy of nearest approach to the Fermi level as a function of angle, $\theta$, in the Brillouin zone.  The essential result is that around the nodal direction in the non-superconducting state, there exists an arc which is ungapped and thus defines a true Fermi surface.  At angles, $\theta$, smaller than a critical angle, $\theta_{h}$, which defines the end of the Luttinger pocket, states of zero excitation energy cease to exist.  There is instead an effective pseudogap which continues to increase as the antinodal direction is approached at small angles.  In the superconducting state a superconducting gap is observed to form on the ungapped Luttinger arcs and the pseudogap region remains relatively unaffected by the superconducting transition.  This behavior follows directly in YRZ theory wherein the pseudogap leads to Fermi surface reconstruction from a large Fermi liquid (FL) tight binding Fermi surface (TBFS) to Luttinger pockets.  These pockets are centered about the nodal direction and are bounded on one side by the antiferromagnetic Brillouin zone boundary.  However, only the side of the Luttinger pocket furthest from the AFBZ boundary is strongly weighted and this forms the well known Fermi arcs while the backside has a smaller weight and can be ignored for many purposes.  In fact, it was not until very recently that this part of the Luttinger pocket was detected in ARPES experiments which previously only found the Fermi arc.  \cite{meng:2009}

In view of the many successes of YRZ theory, as applied to the underdoped cuprates, it is of interest to examine more closely a regime which has so far received very little attention from both theory and experiment.  When the pseudogap is large, corresponding to deep underdoping, one has a Fermi arc around $\theta=\pi/4$ which is followed by a gapped contour of nearest approach energy.  When the doping is increased towards $x=x_c$, corresponding to the QCP of the theory, the large TBFS of FL theory must be recovered.  It is the transition region between the hole Fermi pockets, which are characteristic of the approach to the Mott insulating state, and the TBFS of FL theory that we wish to study.  This transition proceeds through the nucleation of small electron pockets bounded by a surface of zero excitation energies which appear first in the antinodal direction and are additional to the Luttinger hole pockets.  There are no zero energy states on the bridge connecting these two pockets, but rather there is only a contour of nearest approach energies on which there is a finite pseudogap.  As doping is increased further, the gapped contour shrinks in size and eventually, both pockets meet, at $x=x_c$, at which point the TBFS is recovered.  Note that the pockets have two sides, but the YRZ model transfers weight between the sides as the pseudogap is reduced, creating a continuous transition to zero pseudogap.  In that limit, the weakly weighted sides define the AFBZ boundary.  It is the detailed study of the evolution of the electronic structure from a large Fermi surface of Fermi liquid theory to hole Luttinger pockets that is the topic of this paper.  We seek here to provide calculations that should serve as a guide to experiments.
Since ARPES provides us, in principle, with a direct probe of the Fermi surface evolution with doping, we will present our results with this probe in mind.  However, the Fermi surface reconstruction, which is most pronounced in the underdoped region just below the QCP will have an effect on other properties as well.  In particular, we note the possibility of a connection with quantum oscillations associated with electron and hole pockets.  Further, it has been shown to be possible \cite{pushp:2009} to extract from the electronic density of states, $N(\omega)$, which is measured in scanning tunneling spectroscopy (STS) information on the partial contribution to $N(\omega)$ of each angle, $\theta$, in the Brillouin zone.  This type of analysis is very similar to ARPES.\cite{borne:2010}

Section \ref{sec:theory} will describe the relevant theoretical framework and give the parameters involved in the YRZ calculations.  Section \ref{sec:arpes} will discuss the electronic structure of the pseudogap onset as it should appear in ARPES and section \ref{sec:em} includes our description of pocket formation in the context of mass renormalization.  Finally, section \ref{sec:conclusions} will contain a summary of our conclusions.

\section{Theory}
\label{sec:theory}

The YRZ model\cite{yrz:2006,yrz:2009} employs, for the coherent piece, a Green's function,
\begin{eqnarray}
G(\boldsymbol{k},\omega,x)=
\sum_{\alpha=\pm}  \frac{{g_t W^{\alpha}_{\boldsymbol{k}}}}{
{\omega-E^{\alpha}_{\boldsymbol{k}}-\Delta_{\rm sc}^{2}/(\omega+E^{\alpha}_{\boldsymbol{k}})}},
\label{eqn:sc}
\end{eqnarray}
which has been formulated from studies of an effective Hamiltonian in an RVB-type, Gutzwiller projected mean field t-J model.  Entering this propagator are the quantities:
\begin{eqnarray}
E_{\boldsymbol{k}}^ \pm  &= \frac{{\xi_{\boldsymbol{k}}  - \xi_{\boldsymbol{k}}^0 }}{2} \pm E_{\boldsymbol{k}}, \nonumber \\
E_{\boldsymbol{k}} &= \sqrt {\tilde{\xi}_{\boldsymbol{k}}^2  + \Delta
  _{\rm pg}^2 },\nonumber \\
\tilde{\xi}_{\boldsymbol{k}}  &= \frac{(\xi_{\boldsymbol{k}}  + \xi_{\boldsymbol{k}}^0 )}{2}, \nonumber \\
W_{\boldsymbol{k}}^ \pm  &= \frac{1}{2} \left(1\pm \frac{\tilde{\xi}_{\boldsymbol{k}}}{E_{\boldsymbol{k}}}\right). \label{eqn:wplusminus}
\end{eqnarray}
The energy dispersion, $\xi_{\boldsymbol{k}} = - 2t(\cos
k_xa  + \cos k_ya ) - 4t^{\prime} \cos k_xa \cos k_ya - 2t''(\cos
2k_xa  + \cos 2k_ya )-\mu_p$, is the third
nearest-neighbor tight-binding energy dispersion, while $\xi_{\boldsymbol{k}}^0  =  - 2t(\cos k_xa  +
\cos k_ya)$ is the first nearest-neighbor term.  These energy dispersions contain doping dependent
coefficients: $t(x)=g_{t}(x)t_{0}+3g_{s}(x)J\chi/8$, $t^{\prime
}(x)=g_{t}(x)t_{0}^{\prime }$, and $t^{\prime\prime
}(x)=g_{t}(x)t_{0}^{\prime \prime }$,  where $g_{t}(x)=\frac{2x}{1+x}$ and $g_{s}(x)=\frac{4}{(1+x)^{2}}$ are the energy renormalizing Gutzwiller factors for the kinetic and spin terms, respectively.  $g_t (x)$ also appears in Eq.~(\ref{eqn:sc}) as a weighting factor for the coherent part of the Green's function which acts to statistically remove or project out doubly occupied states.
%\cite{gutzwiller:1963}
The dispersion, $\xi_{\boldsymbol{k}}$, uses $\mu_p$ as a chemical potential determined by the Luttinger sum rule in the form,
%\cite{luttinger:1960}
\begin{equation}\label{eq:lutt}
1-x=\frac{2}{(2\pi)^2}\int_{G(\vect{k},\omega=0)>0}d^2k,
\end{equation}
which contains the pseudogap in the dispersion of the propagator, Eq.~(\ref{eqn:sc}).  Thus the chemical potential becomes pseudogap dependent.
 Values of
other parameters in the dispersion were taken from Ref.~\cite{yrz:2006} to be: $t^{\prime}_{0}/t_{0}=-0.3$, $t^{\prime\prime}_{0}/t_{0}=0.2$,
$J/t_{0}=1/3$, and $\chi=0.338$. While here for simplicity we will keep the tight binding parameters introduced in the paper of YRZ, they could be changed to agree better with the measured dispersions in a particular material of interest.

In the YRZ model, both the superconducting gap, $\Delta_{\rm sc}$, and
the input pseudogap, $\Delta_{\rm pg}$, have a $d$-wave $k$-space dependence
described by: 
\begin{eqnarray}
\Delta_{\rm sc}=\frac{\Delta_{\rm sc}^{0}(x)}{2}(\cos k_xa -\cos k_ya)\label{eqn:scgap},\\
\Delta_{\rm pg}=\frac{\Delta_{\rm pg}^{0}(x)}{2}(\cos k_xa -\cos k_ya)\label{eqn:pggap},
\end{eqnarray}
where $a$ is the lattice constant.
  $\Delta_{\rm sc}^0(x)$ and $\Delta_{\rm pg}^0(x)$ are described by the well known superconducting dome and pseudogap line respectively, the latter vanishing at $T=0$ at a QCP in this model.  These are given explicitly as
\begin{align}
\Delta_{\rm{sc}}^0(x)&= 0.14t_0(1-82.6(x-0.16)^2), \label{eqn:scdome}\\
\Delta_{\rm{pg}}^0(x)&= 3t_0(0.2-x). \label{eqn:pgline}
\end{align}
We have used the well known empirical expression of Eqn.~(\ref{eqn:scdome}) for the Tc dome with the assumption that twice the gap to Tc ratio is 6.  Recently, Schachinger and Carbotte \cite{schachinger:2010} have solved a BCS equation for gap and critical temperature in the YRZ model and have found that the opening of a pseudogap has the effect of increasing this gap to Tc ratio as the doping is reduced.  This implies that the superconducting gap remains large in underdoped cuprates, consistent with their small coherence length.

From the YRZ Green's function of Eq.~(\ref{eqn:sc}), one can extract the spectral function, $A(\boldsymbol{k},\omega)$, given by 
  \begin{equation}
  A(\boldsymbol{k},\omega)=\sum_{\alpha=\pm} g_tW_{\boldsymbol{k}}^\alpha [(u^\alpha)^2\delta(\omega-E_{\rm s}^\alpha)+(v^\alpha)^2\delta(\omega+E_{\rm s}^\alpha)],%\\
 % B(\boldsymbol{k},\omega)&=\sum_{\alpha=\pm}  g_tW_{\boldsymbol{k}}^\alpha u^\alpha v^\alpha[ \delta(\omega+E_{\rm s}^\alpha) -\delta(\omega-E_{\rm s}^\alpha)],
  \end{equation}
    where
    \begin{eqnarray}
    E^{\alpha}_{\rm s}&= \sqrt{(E_{\boldsymbol{k}}^{ \alpha })^2  +
  \Delta _{\rm sc}^2 },\\
 (u^\alpha)^2 &=\left[\frac{1}{2}\left(1+\frac{E_{\boldsymbol{k}}^\alpha}{E_{\rm s}^\alpha}\right)\right],\\
 (v^\alpha)^2 &=\left[\frac{1}{2}\left(1-\frac{E_{\boldsymbol{k}}^\alpha}{E_{\rm s}^\alpha}\right)\right].\label{eqn:uvs}
  \end{eqnarray}
Here, $E^\alpha_{\rm s}$ represents the pair dispersion and $(u^\alpha)^2$ and $(v^\alpha)^2$ are standard Bogoliubov quasiparticle amplitudes for their respective $\alpha$ energy branches. 
  
%  
%In addition, we will explore modifications to the input pseudogap $k$-space dependence of the form:
%\begin{equation}
%  \Delta_{\rm pg}={\rm sgn}(\cos k_xa -\cos k_ya)  \Delta_{\rm pg}^{0}(x)\left(\frac{|\cos k_xa -\cos k_ya|}{2}\right)^n\label{eqn:pggapn},
%\end{equation}
%  for variations in $n$.  Note that $n=1$ corresponds to the form of Eq.~(\ref{eqn:pggap}), but all choices of $n$ maintain $d_{x^2-y^2}$ symmetry and are similar to different order harmonic terms.
  
  \section{ARPES}\label{sec:arpes}
%Central to the YRZ model is the existence of electron and hole pockets, shown in Fig.~\ref{fig:ew:a} and Fig.~\ref{fig:ew:b}.  Both are shown for $x=0.18$ which is near $x_c$, and thus has only a small pseudogap.  Fig.~\ref{fig:ew:a} displays the $E_{\vect{k}}^+$ energies as a color scale in one quarter of the Brillouin zone.  Near (0,$\pi$) and ($\pi$,0), one can see small shallow pockets of negative energy ('electron pockets') surrounded by an $E^+_{\vect{k}}=0$ Fermi surface.  Conversely, in Fig.~\ref{fig:ew:b}, the values of $E^-_{\vect{k}}$ are mostly negative in value save for the central hole pocket feature, which is surrounded by an $E^-=0$ Fermi surface.  These Fermi surfaces are the Luttinger surfaces which enter the Luttinger sum rule of Eq.~(\ref{eq:lutt}).
Central to the YRZ model is the existence of electron and hole pockets.
These regions of occupied or unoccupied states overlap in the YRZ model with weighting factors of $W^\pm$ which enter much the same as in BCS theory (Bogoliubov $u^2$ and $v^2$ hole-particle weightings), but here, $\Delta_{\rm sc}$ is replaced with $\Delta_{\rm pg}$ and $\tilde{\xi}_{\boldsymbol{k}}$ replaces the quasiparticle energy as in Eq.~(\ref{eqn:wplusminus}).
While $u^2$ and $v^2$ are centered about the Fermi level, the $W^+$ and $W^-$ are centered about the zeros of a different energy, $(\xi_{\vect{k}}+\xi_{\vect{k}}^0)$, whose zeros acts as the origin of the pseudogap.  This origin corresponds to the TBFS, given by $\xi_{\vect{k}}=0$, being shifted by an effective particle kinetic energy,\cite{anderson:2004} which near half filling corresponds to the AFBZ.  
In the $\Delta_{\rm pg}\rightarrow 0$ limit we therefore establish two boundary energies in this system, that of the TBFS, and that of the AFBZ, wherein $W^\pm$ have values dependent upon the sign of $\xi_{\boldsymbol{k}}+\xi^0_{\boldsymbol{k}}$.
In this limit the TBFS has a weight ($W^\pm$) of $1$ and the AFBZ a weight of zero; thus it is the weighting factors which provide the agency whereby the AFBZ drops out of the quasiparticle energy in this limit, as we know it must.

\begin{figure}
  \centering
  \includegraphics[width=0.5\linewidth]{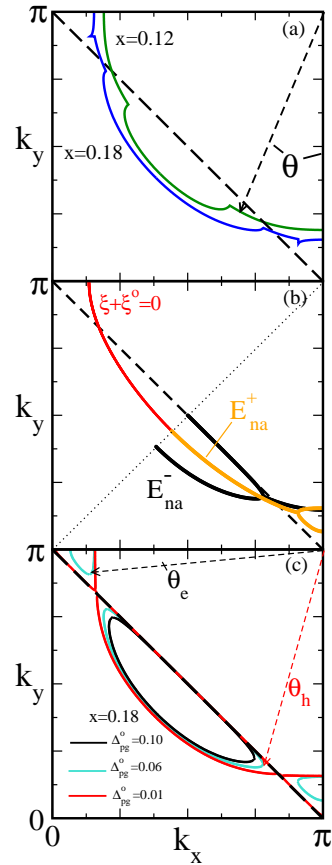}%1x3x18andpgvary-side-raster.eps}
% \subfigure{ \includegraphics[width=0.45\linewidth]{pathx18.eps} }\\
%  \subfigure{\includegraphics[width=0.45\linewidth]{pgvary-nowing.eps}}      
  \caption{\label{fig:contour}(a) Example of heavily weighted nearest approach contours for $x=0.12$ and $x=0.18$.  The influence of the $E^+$ band can be seen in the x=0.18 case near ($\pi$,0) and (0,$\pi$) as the appearance of electron pockets.  (b)  Full nearest approach contours for $E^+$ and $E^-$ in the lower right portion as well as the zeroes of $\tilde{\xi}_{\boldsymbol{k}}$ in the upper left portion.  (c)  Nearest approach pockets for $E^+$ and $E^-$ (here equivalent to Luttinger surfaces) for a single doping of $x$=0.18 for varied input $\Delta_{\rm pg}^0$.  The pseudogap,  $\Delta_{\rm pg}^o=0.06t_0$, corresponds to YRZ fitted parameters for $x=0.18$ as described in the text.}
\end{figure}

A fundamental aspect of the work presented here is the determination of nearest approach contours and energies.  We define these quantities by marking local minima of the $E^+_{\vect{k}}$ and $E^-_{\vect{k}}$ energies in the Brillouin zone, which we will refer to as $E^\pm_{na}$, and that occur at locations $\vect{k}_{na}$.  Often it is convenient to display the energies as a function of angle, $\theta$, as measured from the corner point at $(\pi,\pi)$.  This allows us to compare to a simple $d$-wave superconducting gap such as $\Delta_{\rm sc}(\theta)\approx \cos(2\theta)$.  The angle and magnitude of $\vect{k}$ defines the momentum in the first Brillouin zone.  For some angles there are multiple local minima due to the hole and electron pockets; but also, there are local minima which bridge the pockets.  These connections, or `bridges', do not appear in other contours such as constant energy surfaces nor in the Luttinger surfaces.  In fact, it is for angles along these bridges that the pseudogap has a finite, effective pseudogap value.  Such a set of contours is shown in Fig.~\ref{fig:contour}.

\begin{figure}
  \centering
  \includegraphics[width=0.85\linewidth]{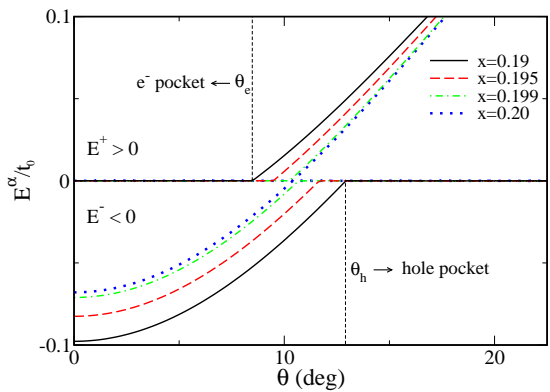}      
  \caption{\label{fig:evstheta}   $E^+$ and $E^-$ energies vs $\theta$ along their individual contours of nearest approach.  Shown for $x=0.19$, 0.195, 0.199 and 0.20 }
\end{figure}

On the $E^+_{na}$ energies, which are $\geq 0$ for all angles, there exists a zero energy Fermi surface for angles $\theta < \theta_e$, which surrounds a region of filled electron states (the `electron pocket') near (0,$\pi$) and ($\pi$,0) and is shown along with the nearest approach contours in Fig.~\ref{fig:evstheta}.
For the $E^-_{na}$ energies, which are $\leq 0$ for all angles, there exists a zero energy Fermi surface for angles $\theta>\theta_h$, which surrounds the region of unoccupied states (the `hole' or Fermi pocket) along the AFBZ boundary.  Important evidence of a Fermi pocket has been obtained experimentally by Meng et al.\cite{meng:2009} on Bi$_2$Sr$_{2-x}$La$_x$CuO$_{6+\delta}$, which revealed not only the presence of a Fermi pocket but also its coexistence  with an additional gapped surface extending the Fermi arc beyond the corners of the Fermi hole pocket.  This experimental observation is an expected feature of the YRZ model as the manifestation of these bridge contours in ARPES data.  

We depict this behavior in Fig.~\ref{fig:contour}(a) where we have shown the most strongly weighted nearest approach contour with energy $\leq 0$.  For angles near 45$^\circ$ we display the back of the Fermi pocket.  For angles $\theta<\theta_h$, which marks the corner of the hole pocket, we display the contour of nearest approach for the $E^-$ band, which has negative energies, as shown in Fig.~\ref{fig:evstheta}.  In this region, the nearest approach energy is a local minima only, caused in the YRZ model by proximity to a zero in $(\xi_{\vect{k}}+\xi^0_{\vect{k}})$ which corresponds to a zero valued minima in $\tilde{\xi}_{\boldsymbol{k}}^2$.  This choice of contour is maintained until $\theta <\theta_e$ at which point the zero energy contour of the $E^+$ band (shown in Fig.~\ref{fig:evstheta}) becomes the more strongly weighted contour of nearest approach.  One will notice that the contour may have kinks when crossing either of $\theta_e$ or $\theta_h$.

The choice of contours in Fig.~\ref{fig:contour}(a) is justified on the criteria that the energies are:
1) a local minima in the absolute value of energy,
2) of negative (or zero) energy,
3) have the largest weighting $W^+$ or $W^-$.  
The full set of contours, for $x$=0.18, as well as the $(\xi_{\vect{k}}+\xi^0_{\vect{k}})=0$ surface are shown in Fig.~\ref{fig:contour}(b).  It is precisely this effective kinetic energy contour given by $\tilde{\xi}_{\boldsymbol{k}}=0$ which forms the bridged regions between the pockets and therefore dominates in the well underdoped, when the hole pockets have become very small, and behaves as an effective Fermi surface in that case.

It can be seen in Fig.~\ref{fig:contour}(c) that changing the magnitude of the pseudogap directly modifies the size of the Fermi pocket.  However, there are two `hard' barriers in the structure of the hole pocket.  The first is the AFBZ boundary.  The second is, in fact, the normal state Fermi liquid TBFS.  As the pseudogap value decreases, the pocket attempts to stretch out to merge with the TBFS.  However, for small pseudogap, the hole pocket is restricted to the bounded region created by the AFBZ and the TBFS.  Further expansion in area is not allowed, as the Luttinger sum rule dictates, in essence, a conservation of the ratio of hole/electron pocket areas.  Therefore, beyond some pseudogap value, for a given doping, electron pockets must emerge in the corner regions (which are also bound by the TBFS and AFBZ) which extend into the upper antiferromagnetic Brillouin zone.  The area of these electron pockets balances the increased area of the hole pocket, finally merging at the intersection of the AFBZ and TBFS for $\Delta_{\rm pg}\rightarrow 0$.
We note that the heavily weighted  hole and electron pockets for $\Delta_{\rm pg}^0=0.06$  evolve into the solid red contours for $\Delta_{\rm pg}^0=0.01$ with the lightly weighted sides collapsing onto the antiferromagnetic Brillouin zone.  In the limit of $\Delta_{\rm pg}^0=0$ this part has zero weight.  This shows how the energy $\xi_{\vect{k}}^0$, which plays a role in YRZ, drops out of consideration in the dispersion curves in the limit of the Fermi liquid, where only $\xi_{\vect{k}}$ plays a role.

\begin{figure}
  \centering
  \includegraphics[width=0.8\linewidth]{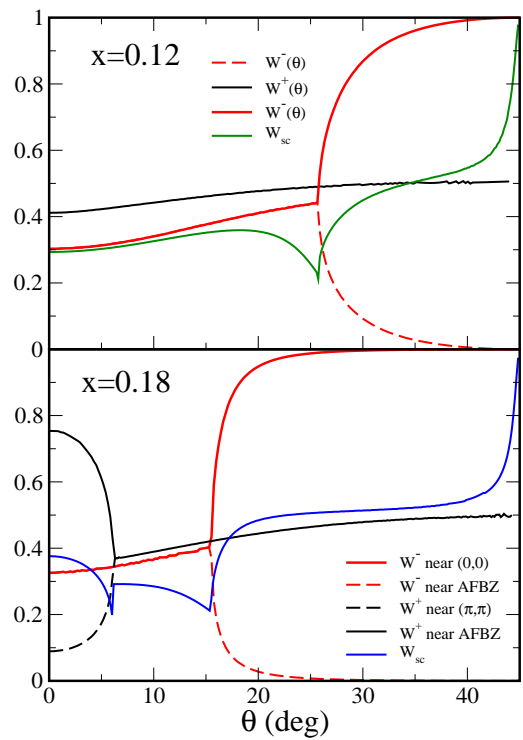}  
 % \subfigure{\includegraphics[width=0.45\linewidth]{x12-weights.eps}  }
  %\subfigure{\includegraphics[width=0.45\linewidth]{x18weights.eps}           }
  \caption{\label{fig:weights}Weighting factors along nearest approach contours.  For the cases with no superconductivity, the weightings are $W^+$ and $W^-$.  The inclusion of superconductivity adds an additional factor of $(v^-)^2$ which represents the occupied state weighting of the Bogoliubov quasiparticles. }
\end{figure}
Examples of the weighting factors along nearest approach contours are shown in Fig~\ref{fig:weights}.  The top frame is for x=0.12, where there are no electron pockets, and the bottom frame is for x=0.18 which lies between $x_{onset}$ and $x_c$ and thus has electron pockets.  
For the x=0.12 case, $W^+_{na}$ is single valued, while for x=0.18, the presence of an electron pocket splits the contour at small angles.  The dashed black line, $\theta<\theta_e$, represents the side of the electron pocket nearest to the AFBZ.  The $W^-_{na}$ weightings are double valued for both dopings and are shown in red, where the dashed curve is now representative of the side of the hole pocket nearest to the AFBZ. It is only at $\theta=\pi/4$ that the solid red line reaches its Fermi liquid value of exactly one.  This occurs even though there exists states of zero excitation energy on the entire hole Luttinger Fermi surface just as in an ordinary Fermi liquid.  Here, $W^\pm<1$ represent a departure from FL theory.  This departure increases particularly quickly as the end of the hole pocket is approached where $W^-(\theta)$ has already dropped below $1/2$ in value for both dopings shown.  
To understand how this comes about, we note that on the hole Luttinger surface $E^-_{\vect{k}_{na}}=0$ and that the weighting factor $W^-_{na}$ on this contour simplifies to
\begin{align}
W^-_{\vect{k}_{na}}&=\frac{1}{2}\left[1-\frac{\tilde{\xi}_{\vect{k}}}{\sqrt{\tilde{\xi}_{\vect{k}}+\Delta_{\rm pg}^2(\vect{k})}}\right]\nonumber\\&=\frac{(\xi_{na}^0)^2}{(\xi_{na}^0)^2+\Delta_{\rm pg}^2(\vect{k}_{na})}.\label{eqn:simpleW}
\end{align}
On the front part of the hole Luttinger pocket $\vect{k}_{na}$ is away from the antiferromagnetic zone boundary, which is defined by $\xi_{\vect{k}}^0=0$, but on the opposing side, we are close to the AFBZ boundary and so we can expect $\xi_{\vect{k}_{na}}^0$ to be small. This means that on the front side $W^-_{\vect{k}_{na}}$ of Eq.~\ref{eqn:simpleW} is of order 1, but is reduced due to the presence of the gap, $\Delta_{\rm pg}(\vect{k}_{na})$, which is squared in the denominator.  Right at $\theta=\pi/4$ the value of $W^-_{\vect{k}_{na}}$ reaches exactly one as shown in Fig.~\ref{fig:weights} (solid red curve) and $W^-_{\vect{k}_{na}}$ has dropped to a little less than $1/2$ its value for $\theta=\theta_h$, marking the end of the hole pocket.  On the other hand, the weight on the backside (dashed red curve) rapidly drops from its maximum at $\theta=\theta_h$ as $\theta$ is increased towards the nodal direction.  As the nodal direction is approached, the contour $\vect{k}_{na}$ shifts closer and closer to the AFBZ boundary and $\xi^0(\vect{k}_{na})$ becomes smaller and smaller and so $W^-_{\vect{k}_{na}}$ of Eq.~\ref{eqn:simpleW} rapidly drops to a value near zero.  On the gapped contour beyond $\theta_h$ for $x=0.12$ (top frame) and in the bridging region between $\theta_h$ and $\theta_e$ for $x=0.18$ (bottom frame) the last equality in Eq.~\ref{eqn:simpleW} no longer holds, but numerical evaluation shows that $W^-_{\vect{k}_{na}}$ continues to drop, reaching a value $\approx 0.3$ in both cases.  These behaviors are quite distinct from the Fermi liquid case of a large Fermi surface for which the weight would remain $1$ at all angles.  

\begin{figure}
  \centering
  	\includegraphics[ width=0.95\linewidth]{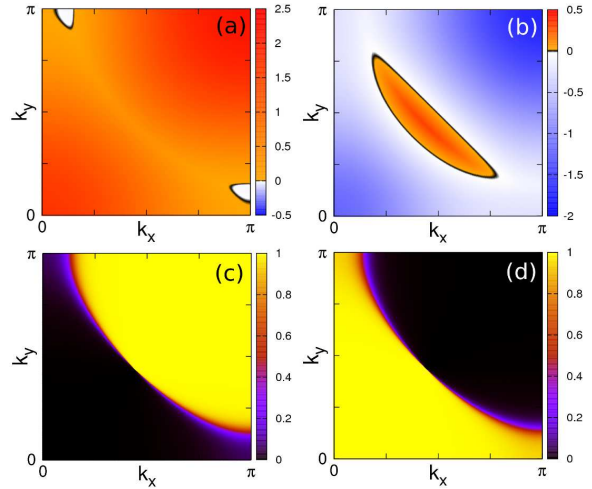}
%  \subfigure{\includegraphics[clip,trim= 27mm 52mm 18mm 48mm,angle=-90, width=0.38\linewidth]{n1-x18-eplus-new.eps}\label{fig:ew:a}}                
%  \subfigure{\includegraphics[clip,trim= 27mm 52mm 18mm 48mm,angle=-90, width=0.38\linewidth]{n1-x18-eminus-new.eps}\label{fig:ew:b}} 
%    \subfigure{\label{fig:uddd}\includegraphics[clip,trim= 27mm 52mm 18mm 48mm,angle=-90, width=0.38\linewidth]{n1-x18-WPlus-new.eps}\label{fig:ew:c}}                
%  \subfigure{\label{fig:opddt}\includegraphics[clip,trim= 27mm 52mm 18mm 48mm,angle=-90, width=0.38\linewidth]{n1-x18-WMinus-new.eps}\label{fig:ew:d}} 
  \caption{\label{fig:ew}One quarter of Brillouin zone displaying: Top Row: $E^+_{\vect{k}}$ and $E^-_{\vect{k}}$ in (a) and (b) respectively, for the $x$=0.18 case where the color scale has units of $t_0$. Bottom Row: $W^+_{\vect{k}}$ and $W^-_{\vect{k}}$, (c) and (d) respectively, also for the x=0.18 case.}
\end{figure}

\begin{figure}
  \centering
  \includegraphics[width=0.85\linewidth]{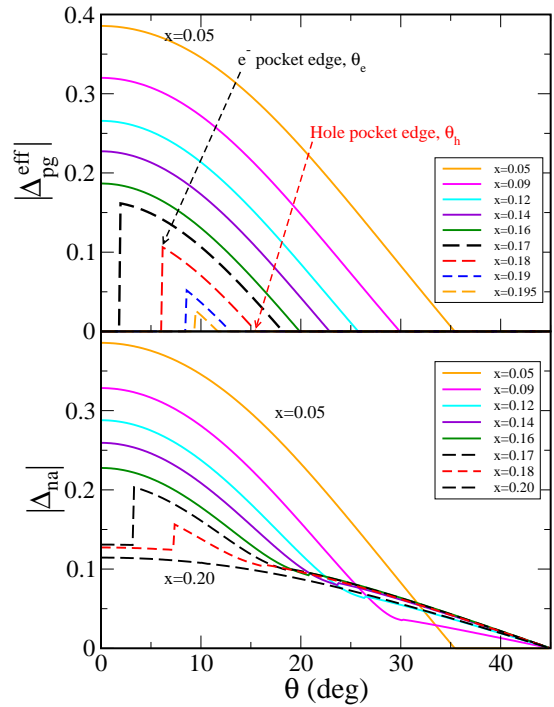}
  %\subfigure{ \includegraphics[width=0.45\linewidth]{pgeffective.eps} }\\
  %\subfigure{\includegraphics[width=0.45\linewidth]{pgeffective-sc.eps}          }
 \caption{\label{fig:pgeffective}Top frame: $\Delta_{\rm pg}^{eff}$ are the nearest approach energies of the most strongly weighted of $E^-$ or $E^+$ bands, for various dopings.  For large pseudogap, the system is dominated by the $E^-$ band.  At $x$=0.17 the $E^+$ band provides an $E=0$ Fermi surface at small angles with a strong weighting.  Bottom Frame:  Same as above but for $E_{\rm sc}^\pm$.  This gives an additional contribution from $\Delta_{\rm sc}$.  Note for comparison to ARPES plots, we have taken the absolute value of these energies.}
\end{figure}

The behavior of $W^+_{\vect{k}_{na}}$ associated with the nearest approach contour of the $E^+_{\vect{k}}$ (solid black line) is most interesting for $x=0.18$ (bottom frame).  In this case, for $\theta$ between $0$ and $\theta_e$ we have two more zero energy contours; one with large weight (solid line) and one with smaller weight (dashed line).  The weights on these surfaces, however, remain intermediate, well above $0$ and well below $1$.  It is in this respect that these electron pockets maintain continuous electronic states which support closed orbits within a single band unlike the hole pocket, which has no continuous state in the nodal direction near the AFBZ.

%Beyond the critical angle, $\theta_h$, the weight $W^-(\theta)$ continues to drop in magnitude and for $x=0.12$ it reaches $\approx 0.3$ at the end of the bridge contour of nearest approach.  By contrast, the weight $W^+(\theta)$ (solid black) starts off below one ($\approx 1/2$) at the largest angles shown and drops slightly for $x=0.12$ as the antinodal direction is approached.  For $x=0.18$, which is closer to the QCP, a second pocket forms (electron pocket) and $W^+(\theta)$ becomes double valued for small angles, but remains well below one, even on the strongly weighted part of the electron pocket (\textbf{maybe relate this to increased effective mass in this region}).  

There are complications involved in the inclusion of superconductivity.  In this case we must also include the Bogoliubov $v^2_{\vect{k}}$ weightings,
which take the form of Eq.~(\ref{eqn:uvs})
   We show the $W^+_{na}{v^+_{na}}^2$ (solid green curve in Top frame of Fig.~\ref{fig:weights}) and $W^-_{na}{v^-_{na}}^2$ (solid blue curve in bottom frame of Fig.~\ref{fig:weights}), along the contours shown in Fig.~\ref{fig:contour}(a).
 On the surfaces defining the hole and electron pockets, $E^-_{\vect{k}}$ and $E^+_{\vect{k}}$ are respectively zero, so that on both the front and back side of these surfaces, the Bogoliubov ${v^-}^2$ is precisely $1/2$.  This is shown as the green and blue solid lines in Fig.~\ref{fig:contour}(a).  For $\theta<\theta_h$ and $\theta$ in the interval $\theta_h$ to $\theta_e$ for $x=0.12$ and $x=0.18$ respectively, the reduction in $W{v^-}^2$ from $W$ alone is less than $1/2$ because on these contours $E^-_{\vect{k}_{na}}$ is negative and finite.  Hence, ${v^-}^2$ in Eq.~(\ref{eqn:uvs}) is greater than $1/2$.  For the specific case of $x=0.12$ (top frame) the solid red and solid green contours have practically merged at $\theta=0$ while in the lower frame, the solid blue contour increases towards the solid red until $\theta=\theta_e$ at which point we change from the bridging contour to the new electron surface associated with the electron pockets.
  
  %  This illustrates that the weightings of these contours in the superconducting state are much more constant than in the normal pseudogapped state.  Note that on the Luttinger contours, which correspond to zero energy in the non-superconducting case, the weighting in the superconducting state is exactly half its normal pseudogap state value, but on the bridging contour, the relationship is more complicated.  A prediction of YRZ theory which is \textbf{NOTE: worth pointing out is that for $x=0.12$, the superconducting and non-superconducting weights, $W^-$ are nearly the same for $\theta$ small, as is shown in Fig. 4a}
For completeness, we illustrate the $E^\pm$ energies and their weighting factors in the full Brillouin zone in Fig.~\ref{fig:ew}.  The electron pockets are shown in white near the antinodes of Fig.~\ref{fig:ew}(a), while the hole pocket is located in the nodal region and is shown in orange in Fig.~\ref{fig:ew}(b).  The spectral functions of the two dispersions are overlaid with their respective weighting factors shown in Fig.~\ref{fig:ew}(c)and (d).  It should be clear that the weighting of the $E^-$ bridges should be reduced as compared to the Fermi pocket itself.

Recent experimental determination of gap values in Bi-2201 using ARPES \cite{kondo:2009} have been qualitatively described by the YRZ model for the well underdoped case where the pseudogap dominates the system \cite{leblanc:2010}.  Here we present similar calculations over a full doping range and demonstrate the additional features created by electron pockets in the small pseudogap case.
Fig.~\ref{fig:pgeffective}(a) shows this effect in the absence of superconductivity.  For $x=0.20$ we would obtain a full zero energy, gapless Fermi surface.  For small dopings, say $x=0.05$, we have a strong pseudogap which follows a modified $\cos(2\theta)$ function, terminating at $\theta_h \approx 35^\circ$.  
This qualitative structure is maintained until the doping $x_{onset}$ where the electron pocket has now formed a zero energy TBFS with strong weighting at small angles, $\theta<\theta_e$, similar to the $E^+_{na}$ energies shown in Fig.~\ref{fig:evstheta}.
Therefore, for dopings where $x_{onset}<x<x_c$, the curve appears as a modified cosine function for angles $\theta>\theta_e$, and drops to zero for $\theta<\theta_e$.  This leads to the `russian doll' type arrangement of dashed curves covering the four largest values of $x$ considered in the figure.  This feature is still present when including a superconducting gap, as is shown in Fig.~\ref{fig:pgeffective}(b).  The extreme cases of $\Delta_{\rm sc}=0$ or $\Delta_{\rm pg}=0$ produce what can be approximated as cosine functions terminating at $\theta_h$ and 45$^\circ$ respectively.  Again, there is a narrow region of doping between $x_{onset}$ and $x_c$ wherein the behaviors of these curves are modified.

For $\theta<\theta_e$ and $\theta>\theta_h$, the behavior is that of the superconducting gap.  This corresponds to the contour of the normal state Fermi surface where the superconducting gap is dominant.  For $\theta_e<\theta<\theta_h$ the behavior is that of the finite valued, but negative, $E^-$ band energy which is interpreted as an effective pseudogap value.  This corresponds to the pseudogapped out region between the hole and electron pockets which follows along the nearest approach bridges where there is no Fermi surface in the normal pseudogapped state (see Fig.~\ref{fig:contour}(b)).  

\begin{figure}
  \centering
  \includegraphics[width=0.85\linewidth]{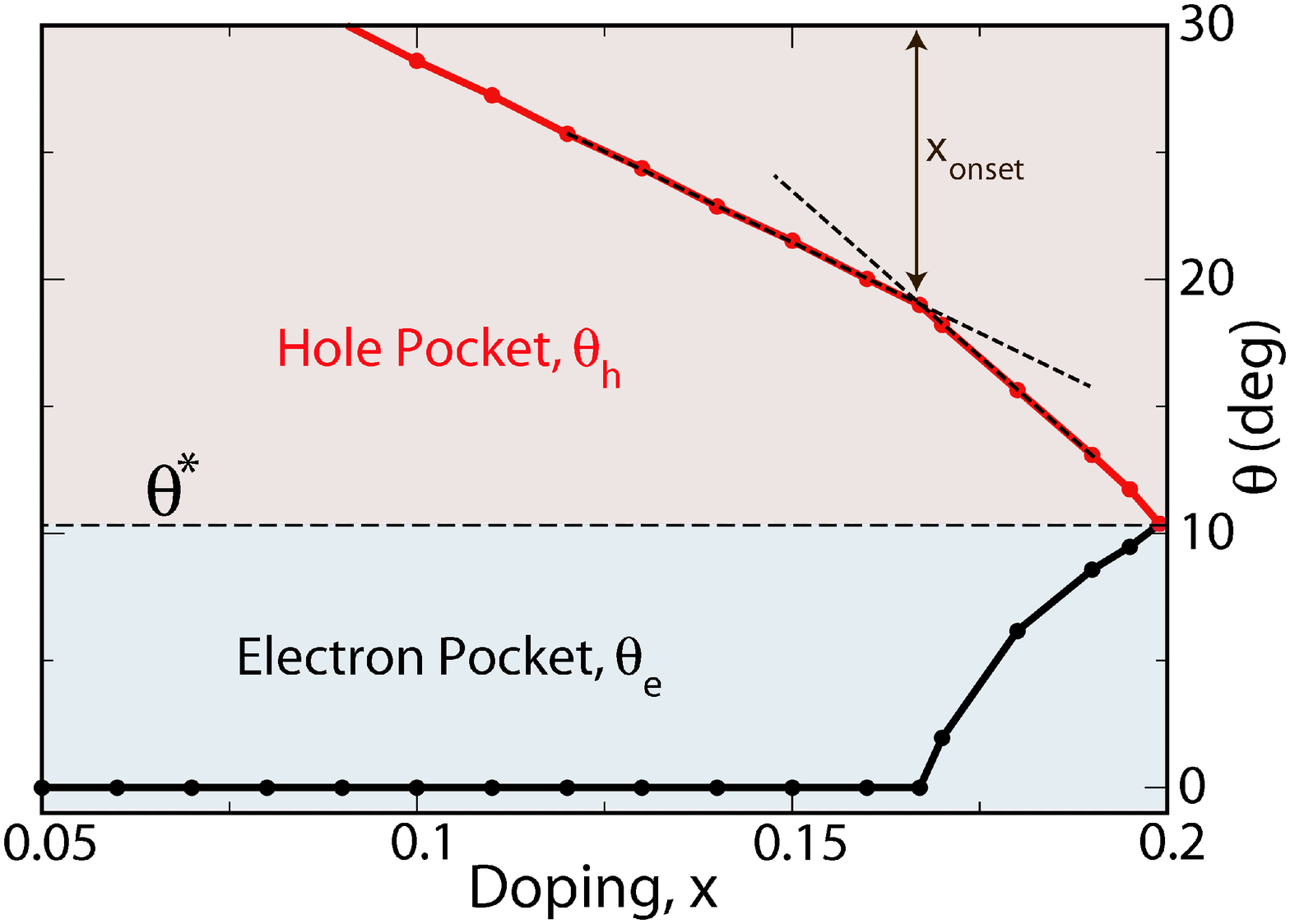}          
  \caption{\label{fig:thetacrit}Progression of hole and electron pockets, defined by their angles of onset, $\theta_e$ and $\theta_h$ as a function of doping converging to $\theta^*\approx 10^\circ$ at $x=x_c=0.20$.}
\end{figure}

Of primary interest for this work is the behavior of the pocket edges.  Fig.~\ref{fig:thetacrit} depicts the calculated $\theta_e$ and $\theta_h$ critical angles across all dopings and their convergence to the value $\theta^*$ at $x=0.20$.  This figure is separated into two regions for $\theta<\theta^*$ and $\theta>\theta^*$ for the electron and hole pocket angles respectively.  The angle $\theta_e$, which is always less than $\theta^*$, maintains a zero value corresponding to the absence of an electron pocket until $x=x_{onset}$ above which $\theta_e$ is finite, while  $\theta_h$, which is always greater than $\theta^*$, has finite value at all relevant dopings.  The angles $\theta_h$ and $\theta_e$ converge at $x=x_c=0.20$ to a value of $\theta\approx10^\circ$, which corresponds to the intersection of the TBFS and the AFBZ.  The doping dependence of $\theta_h$ is nearly linear away from $x_{onset}$, showing two distinct regions, $x<x_{onset}$ and $x>x_{onset}$, which differ in slope as illustrated by the black dashed lines which intersect at $x_{onset}$.  The modification to the slope coincides with the onset of electron pockets.  Experimental verification of the modification of this slope would be evidence of the presence of shallow electron pockets forming as dictated by the Luttinger sum rule.  We hope our work will stimulate such experiments.

%\begin{figure}
%  \centering
%  \includegraphics[width=0.65\linewidth]{muvsn.eps}          
%  \caption{\label{fig:chem}Chemical potential, $\mu_p$ as a function of n, as determined by the Luttinger sum rule.}
%\end{figure}

\section{Effective Mass}\label{sec:em}
We wish to discuss the effective electronic mass on the Luttinger surfaces in the pseudogapped state.
In the normal pseudogap state, one can reformulate Eq.~\ref{eqn:sc} as it was presented in Ref.~\cite{yrz:2006} by using the pseudogap self energy which is given by: $\Sigma_{\rm pg}(\vect{k},\omega)=\Delta_{\rm pg}^2(\vect{k})/[\omega+\xi^0_{\vect{k}}]$, which depends on momentum $\vect{k}$ as well as on energy, $\omega$.  
The inclusion of a pseudogap self energy acts to directly renormalize the energy of the quasiparticles.  These renormalized quasiparticles will behave as if having a new renormalized effective mass, which we now calculate.
To begin this, for $\omega\rightarrow 0$, the Green's function has a simple pole at
\begin{equation}
\omega=\left[\xi_{\vect{k}}+\frac{\Delta^2_{\rm pg}(\vect{k})}{\xi^0_{\vect{k}}}\right]\left[1+\frac{\Delta^2_{\rm pg}(\vect{k})}{(\xi^0_{\vect{k}})^2}\right]^{-1}\label{eqn:omega}.
\end{equation}
  We evaluate Eq.~(\ref{eqn:omega}) on the Luttinger surface which defines a contour in momentum space $\vect{k}\equiv[k_F,\theta]$ where $k_F$ is the magnitude of $\vect{k}$ at fixed angle $\theta$.    In the case of multiple $k_F$ for a given $\theta$, we are interested in only the strongly weighted portion, as displayed in Fig.~\ref{fig:weights}.  However, on this contour, $E^-_{\vect{k}}$ of Eq.~(\ref{eqn:wplusminus}) is zero and therefore
\begin{equation}
-\xi_{k_F,\theta}\xi^0_{k_F,\theta}=\Delta_{\rm pg}^2(k_F,\theta)\label{eqn:b}
\end{equation}
For $k$ near $k_F$ at fixed $\theta$ we use $k=k_F+\delta k$. As a result, $\omega$ in Eq.~(\ref{eqn:omega}) can be rewritten as 
\begin{equation}
\omega=\Bigg(\frac{d}{dk}\left\{\xi_{\vect{k}}+\frac{\Delta^2_{\rm pg}(\vect{k})}{\xi^0_{\vect{k}}}\right\}\left[1+\frac{\Delta^2_{\rm pg}(\vect{k})}{(\xi^0_{\vect{k}})^2}\right]^{-1}\Bigg)_{\vect{k}=k_F,\theta}\delta k,\label{eqn:diffomega}
\end{equation}
where use was made of Eq.~(\ref{eqn:b}).  

\begin{figure}
  \centering
  \includegraphics[width=0.85\linewidth]{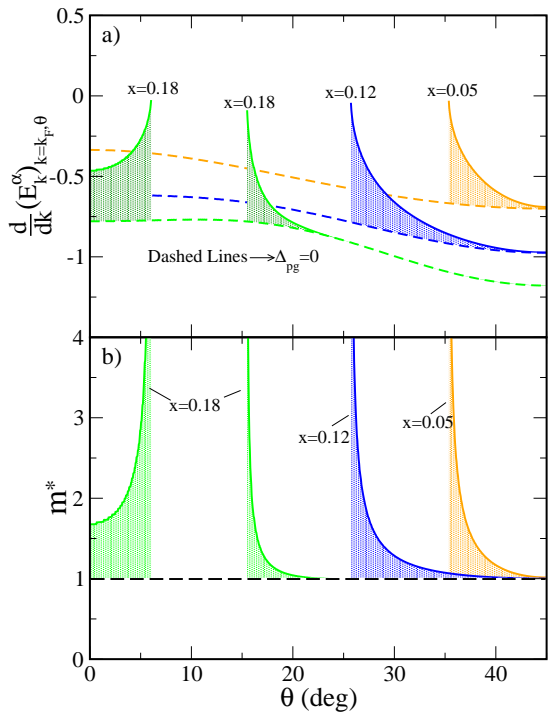}
   \caption{\label{fig:massrenorm} a) The slope of $E^\alpha_{\vect{k}_{na}}$ crossing $k=k_F$ as a function of $\theta$.  The solid curves have pseudogap, while the dashed curves are the slopes across the underlying TBFS of $\xi(\vect{k})$.  Here $\alpha$ is determined as in Fig.~\ref{fig:ew}(a), as the energy branch that maintains coherent quasiparticles.  b) The ratio of slopes in (a), $\frac{d}{dk}[E^\alpha_{\vect{k}_{na}}(\Delta_{\rm pg})]\Big/\frac{d}{dk}[E^\alpha_{\vect{k}_{na}}(\Delta_{\rm pg}=0)]$, which is also given precisely by the effective mass ratio of Eqn.~\ref{eqn:messy}. }
\end{figure}

In the underlying Fermi liquid state, where $\Delta_{\rm pg}^2(\vect{k})=0$, the corresponding quasiparticle pole with which we wish to compare would be at $\omega=\frac{d\xi_{\vect{k}}}{dk}\Big|_{\vect{k}=k_F,\theta}\delta k$.  The ratio of these poles gives the ratio of renormalized to unrenormalized electronic mass coming from the pseudogap formation as:
\begin{align}
m^*_{\theta}&=\frac{m^*_\theta(\Delta_{\rm pg})}{m^*_\theta(\Delta_{\rm pg}=0)}\nonumber\\&=\left[1+\frac{\Delta_{\rm pg}^2(k_F,\theta)}{(\xi^0_{k_F,\theta})^2}\right]\left[1+\frac{\frac{d}{dk}\left(\frac{\Delta_{\rm pg}^2(\vect{k})}{\xi^0_{\vect{k}}}\right)}{\frac{d\xi_{\vect{k}}}{dk}}\right]^{-1}_{\vect{k}=k_F,\theta}.\label{eqn:messy}
\end{align}
Here we have made the approximation that the underlying Fermi liquid band structure varies only slowly with $\vect{k}$ around the Fermi energy and have evaluated $\xi_{\vect{k}}$ at the pseudogap nearest approach contour rather than the Fermi liquid contour.  This is expected to be a good approximation as the two contours do not differ much on the strongly weighted side.  Fundamentally, we are restricted from evaluating this quantity in this simple manner on the opposing lightly weighted side of the pocket as the regions in k-space differ completely for the two cases, with and without pseudogap, causing the approximations here to break down.  Results for the effective mass ratio of Eq.~(\ref{eqn:messy}) are presented in the lower frame of Fig.~\ref{fig:massrenorm} for three values of doping.  Before describing these in detail, we note two limits of Eq.~(\ref{eqn:messy}).  First, in the nodal direction, $\Delta_{\rm pg}(\vect{k})=0$ and all energies reduce to the Fermi liquid unrenormalized band structure, in which case $m^*_\theta(\Delta_{\rm pg})/m^*_\theta(\Delta_{\rm pg}=0)=1$.  Also, when the hole pocket ends at $\theta=\theta_h$ the derivative of $E^-_{\vect{k}}$ on the Luttinger contour with respect to $|\vect{k}|$ at fixed $\theta$ vanishes, ie. $\frac{d}{dk}E^-(k,\theta_h)=0$ for fixed $\theta=\theta_h$.  This leads to the relation
\begin{equation}
\frac{d}{dk}\left[\xi_{\vect{k}}\xi^0_{\vect{k}}+\Delta_{\rm pg}^2(\vect{k})\right]=0\label{eqn:e}
\end{equation}
from which we can conclude, making use of Eq.~(\ref{eqn:b}), that the quantity in the denominator of Eq.~(\ref{eqn:messy}) vanishes and so the effective mass renormalization is infinite.  The quasiparticles at the termination of the Luttinger hole pocket have become infinitely sluggish to continued progression in $\theta$.  To treat the case when electron pockets are present is a straightforward generalization of the above to their respective contours.

In Fig.~\ref{fig:massrenorm} we summarize our numerical results for three values of doping $x=0.05$ (heavily underdoped) $x=0.12$ and $x=0.18$ (which is near our chosen quantum critical point at $x=x_c=0.2$). In the top frame we show the slopes of the electron dispersion curves across the Luttinger contour at $\omega=0$.  The Fermi liquid case corresponds to the dashed curves and the renormalized YRZ pseudogap case corresponds to the solid curves.  The shaded region helps to see at a glance the difference between these two cases.  As we have already discussed, at $\theta=\pi/4$ (nodal direction) there is no slope renormalization.  Away from this point, however, the mass renormalization 
%(lower frame of Fig.~\ref{fig:massrenorm}) 
increases and at the end point of the Luttinger pocket it diverges as we have already demonstrated on the basis of Eqns.~(\ref{eqn:messy})  and (\ref{eqn:e}).  For any doping exhibiting electron pockets ($x_{onset} < x< x_c$) there is an additional region about the antinodal direction which starts at $\theta=\theta_e$ and ends at $\theta=0$.  At $\theta=\theta_e$ the mass is again infinite, as we have zero slope for the quasiparticle dispersions after which it gradually decreases and has its minimum of about 1.7 at $\theta=0$, in the case shown ($x=0.18$), normalized to the Fermi liquid case.  The value of $m^*(\theta=0)$ will shift continuously to $1$ as the pseudogap is reduced in magnitude or $x$ approaches $x_c$. 
%(bottom frame of Fig.~\ref{fig:massrenorm}). 
 On the other side of $x_{onset}$, the physics of the approach to the Mott insulating state at half filling proceeds with a loss of electron states which have a zero excitation gap.  These reside on the significantly weighted Luttinger arc which decreases in length as doping, $x$, is decreased.  In addition, the quasiparticle effective mass ($m^*$) increases above its Fermi liquid value as the end of the arc is approached where $m^*$ diverges.

In all cases, however, there remain coherent quasiparticles, granted fewer and fewer, as the Mott transition is approached, as is also observed in the work of Lee et.~al.~\cite{lee:2005}  These authors find a coherent optical response which remains down to the lowest dopings considered, well past the end of the superconducting dome in the underdoped cuprates.  A second observation long considered anomalous, which finds a natural but very robust explanation in the YRZ model, is the work of Zhou et. al.\cite{zhou:2003}  who found that the slope of the dispersion curves in the nodal direction in LSCO are only very weakly dependent upon doping.  We see here that this slope remains unaffected by the formation of a pseudogap and hence it is dependent only on the underlying band structure.  The energy scale involved in this case is the bandwidth scale or, more precisely, the effective first nearest neighbor hopping scale ($t_0$).  Modification of doping only involves small changes in the position of the Fermi momentum with respect to its value at half filling, and this will not change the corresponding nodal direction Fermi velocity much, as is observed.  There also exists evidence \cite{kanigel:2006,kanigel:2007,norman:2007} that the length of the ungapped arcs about the nodal direction increases in a roughly linear fashion with temperature.  This could be due to a reduction in pseudogap amplitude with increasing temperature, T, which can be justified with the inclusion of scattering.  It could also be modeled with an input pseudogap which opens more slowly with angle out of the nodal direction as one raises the temperature.

Although not rigorously explored here, we take note of the consistency of the antinodal effective mass results along the small electron pocket in the x=0.18 case.  The value of $m^*\approx1.7 $ at $\theta=0$ agrees well with the results of de Haas-van Alphen (dHvA) and Shubnikov-de Haas (SdH) quantum oscillation experiments.  These experiments have reported values of $m^*_{\rm dHvA}=1.76\pm 0.07$ and $m^*_{\rm SdH}=1.9 \pm 0.1$\cite{jaudet:2008,doiron:2007}, consistent with electron pockets \cite{leboeuf:2007}.  Further, the area of the electron pockets is restricted in YRZ by the underlying TBFS, and AFBZ to a value of at most $2-3\%$ of the Brillouin zone, also in reasonable agreement with experiment.  Of course, the weightings of Fig.~\ref{fig:weights} results in a reduction of coherent states on the hole pocket near AFBZ which could suppress the signal of oscillation from the hole pocket.  By contrast, the weighting along the electron pocket always remains significant in magnitude. There may be conditions under which the hole pocket can sustain coherent oscillations, consistent with a hole area ($A_h$) of percent given by the Luttinger sum rule to be $A_h=\frac{x}{2}+A_e$ where subscripts $e$ represents the emerging electron pocket.  Otherwise, it may be that, with loss of fully weighted pockets, oscillations in YRZ should revert to a Fermi arc picture where it has been shown that quantum oscillations could still arise but with the frequency of oscillation related to the arc length, rather than to the area enclosed by the pockets.\cite{pereg:2010}  Clearly, the full calculation of quantum oscillations in the YRZ model would produce interesting results.

\section{Conclusions}\label{sec:conclusions}

We have studied how the electronic structure in the cuprates evolves due to the opening of a pseudogap in the model of Yang, Rice and Zhang.  The pseudogap is associated with the antiferromagnetic Brillouin zone boundary and has its microscopic origin in a RVB spin liquid.  While we considered the entire underdoped region of the phase diagram, an emphasis has been placed on the region near the quantum critical point at doping $x=x_c$ below which the pseudogap becomes finite.  We also emphasize how the large Fermi surface of Fermi liquid theory progressively transforms into a small Luttinger hole pocket centered about the nodal direction, a characteristic feature of the loss of metallicity on the approach to a Mott insulating state at half filling.  Just below $x_c$ (QCP), important modifications occur at the point of crossing between the TBFS and the AFBZ, where there is no longer a zero energy contour.  This breaks the large Fermi surface into two pieces separated by a bridging region. Both pieces are able to sustain zero energy excitations, but the bridging region, between a large hole pocket, in the nodal direction, and the small electron pocket near the antinodal direction, is gapped.  Each pocket has two sides, one with a weighting factor near one and the other with a smaller weighting closer to zero.  The strongly weighted piece is close to the large TBFS while the weakly weighted piece is close to the AFBZ line associated with half filling which corresponds to a Mott insulating state.  As the doping is further reduced, the hole pocket shrinks in size as does the electron pocket which eventually disappears entirely leaving only a hole pocket (in each quadrant). 
% When the electron pockets exist, however, the energy of nearest approach in a direction $\theta$ which crosses this pocket displays two zeroes (i.e. a real Fermi surface exists with zero excitation energy).  In between these two zeroes we have a shallow negative energy region of occupied states.  One of the zeroes corresponds to a large weighting and the other to a small value.
%  The heavily weighted contour gives the remaining Fermi surface at this angle, $\theta$, 
  When it exists, the electron pocket is limited to angles in the interval from $\theta=0$ to $\theta=\theta_e$ in the antinodal region.   A similar situation holds for the hole pocket
  % Denoting the end of the hole pocket by $\theta_h$, and again retaining only the heavily weighted side of the Luttinger pocket we obtain a second piece of the Fermi surface 
  which extends from $\theta_h$ to $\pi/4$.  
  
The behavior of the effective pseudogap as a function of angle along the bridging contour is particularly interesting.  Its value  starts from zero at $\theta=\theta_h$ and grows until $\theta=\theta_e$ at which point it shows a discontinuous drop to zero as we move to the electron pocket piece of the Fermi surface which is not continuously connected to the bridging contour.
  When superconductivity is present there is a superconducting gap on the ungapped parts of the normal state Fermi contour, but in the region between $\theta_e$ and $\theta_h$, there remains an enhancement, over its superconducting state value, of the effective gap measured in ARPES.  Of course when we approach half filling the electron pocket ceases to exist and $\theta_e=0$.  We find that $\theta_e$ is non-zero only above $x=x_{onset}\approx0.167$,  a value which depends somewhat on the assumptions made for the angular dependence of the input pseudogap that enters the YRZ model, as well as details of the band structure parameters of the model.
  These authors also use the lowest d$_{x^2-y^2}$ harmonic, Eq.~\ref{eqn:pggap}, for the momentum variation of the pseudogap in the Brillouin zone.  If this assumption is relaxed and this harmonic is taken to power $n$ then variations in the exponent, $n$, can change the shape of the Luttinger contours as well as  the onset doping, $x_{onset}$.  These are, however, secondary considerations and the results we have just described remain almost unmodified for large variations in the input pseudogap angular profile; for example, a $\vect{k}$ independent pseudogap.  Thus the physics described here is robust.

On the Luttinger part of the nearest approach contours, the effective pseudogap is zero.  Nevertheless, the energies in this region are modified from those associated with the underlying TBFS.  A measure of these changes is the change in effective mass at fixed angle, $\theta$, that is brought about by the presence of the input pseudogap.
% which is non-zero everywhere except at the points in the Brillouin zone where the pseudogap changes sign in order to respect its $d$-wave symmetry.  
 The effective mass is unchanged from its tight binding value only in the nodal direction. As we move towards the end of the Luttinger hole pocket the effective mass increases and has a divergence right at the corner where we go from linear in momentum, away from the Fermi momentum, to a gapped behavior for the renormalized dispersion curves at fixed angle, $\theta$.  
A similar effect applies when the doping is sufficiently close to the QCP at $x=x_c$ that an electron pocket also exists.  In this case, the renormalized mass is larger than the bare mass everywhere and diverges at $\theta=\theta_e$.

The picture that emerges for low doping, near half filling, is an approach to the Mott transition where the number of ungapped quasiparticles available becomes greatly reduced, residing only on the Luttinger surface around the nodal direction.  In addition, as the size of the Luttinger hole pocket shrinks the effective mass of the few remaining quasiparticles on the heavily weighted Fermi arc retains its tight binding value only at $\theta=\pi/4$ but increases as the end of the Luttinger pocket is approached where there is a divergence.

\begin{acknowledgements}
  We thank E.J. Nicol for many discussions on the physics of the pseudogap model employed here and E. Schachinger for contributions during the initial stages of this work. This research has been supported by the Natural Sciences and
Engineering Research Council of Canada (NSERC) and the Canadian Institute
for Advanced Research (CIFAR).
\end{acknowledgements}
\bibliographystyle{spphys}
\bibliography{bib}

\end{document}